\documentclass[11pt, letterpaper]{article}

\usepackage[margin=1in]{geometry}

\usepackage{amsmath,amssymb}
\usepackage{graphicx}
\usepackage{booktabs}
\usepackage{cite}
\usepackage{hyperref}
\usepackage{url}
\usepackage{lscape}
\usepackage{adjustbox}
\usepackage{float}

\usepackage{mathptmx} 

\usepackage{titlesec}
\titlespacing{\section}{0pt}{*2}{*1}
\titlespacing{\subsection}{0pt}{*1.5}{*0.8}

\title{The Impacts of Increasingly Complex Matchup Models on Baseball Win Probability}

\author{
    Tristan Mott\thanks{Correspondence to: \texttt{tjmott@byu.edu}}, Caleb Bradshaw, David Grimsman, and Christopher Archibald \\[0.5em]
    \small \textit{Computer Science Department} \\
    \small \textit{Brigham Young University}
}

\date{April 2025}

\begin{document}

\maketitle

\begin{abstract}
\noindent
Baseball is a game of strategic decisions including bullpen usage, pinch-hitting
and intentional walks. Managers must adjust their strategies based on the changing state of the game in order to give their team the best chance of winning. In this thesis, we investigate how
\emph{matchup models}---tools that predict the probabilities of plate appearance
outcomes---impact in-game strategy and ultimately affect win probability. We
develop four progressively complex, hierarchical Bayesian models that predict
plate appearance outcomes by combining information from both pitchers and
batters, their handedness, and recent data, along with base running probabilities
calibrated to a player’s base-stealing tendencies.

Using each model within a game-theoretic framework, we approximate
subgame perfect Nash equilibria for in-game decisions, including substitutions
and intentional walks. Simulations of the 2024 MLB postseason show that more
accurate matchup models can yield tangible gains in win probability---as much
as one additional victory per 162-game season. Furthermore, employing the most
detailed model to generate win predictions for actual playoff games demonstrates
alignment with market expectations, underscoring both the power and potential
of advanced matchup modeling for on-field strategy and prediction.

\vspace{1em}
\noindent
\textbf{Keywords:} Baseball analytics, Hierarchical Bayesian modeling,
Game theory, Win probability, MLB simulations
\end{abstract}


\clearpage
\tableofcontents

\clearpage
\section*{LIST OF TABLES AND FIGURES}

\vspace{1\baselineskip}
\noindent
\textbf{TABLES}

\noindent
Table \ref{tab:logloss}. Comparison of log loss and geometric mean probability (GMP) on the validation set, as well as GMP recalculated assuming BR is ground truth. \dotfill \pageref{tab:logloss}\\
Table \ref{tab:ROI}. Monte Carlo-sampled ROIs with various ``cushion'' thresholds when betting against consensus lines. \dotfill \pageref{tab:ROI}\\

\vspace{2\baselineskip}
\noindent
\textbf{FIGURES}

\begin{figure}[htbp]
\centering
\includegraphics[width=1.0\textwidth]{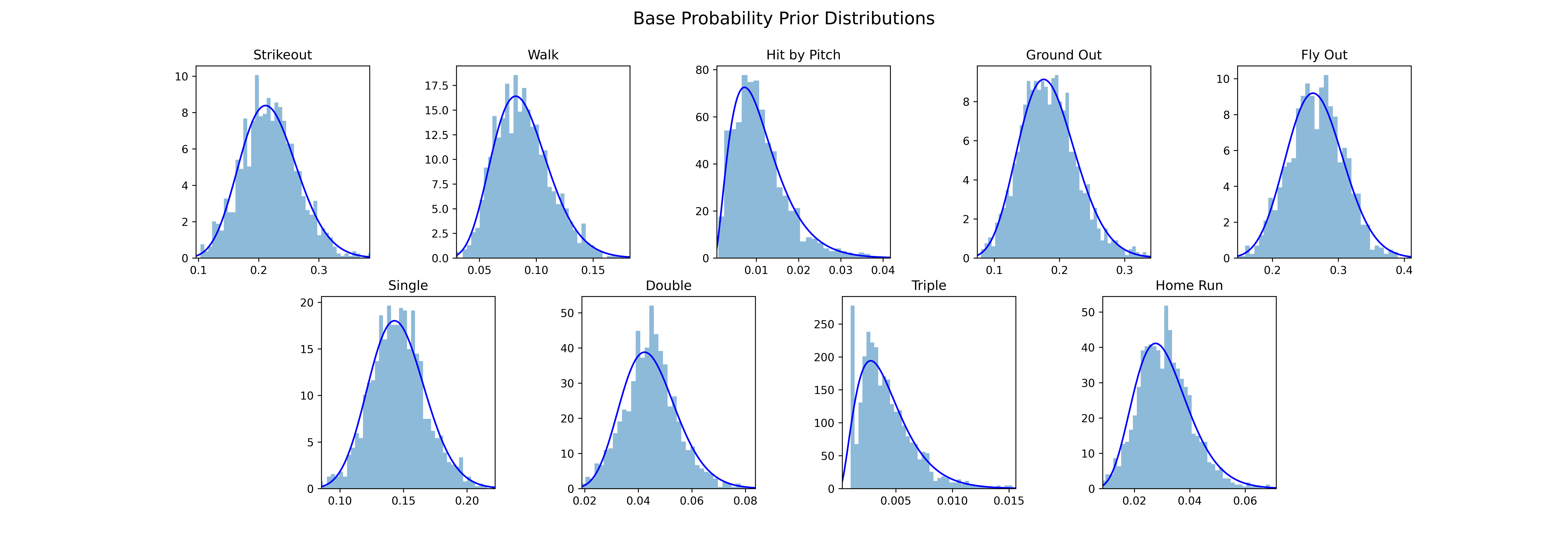}
\caption{Base probability prior distributions for the 9 plate appearance outcomes, estimated from pitchers with more than 100 plate appearances since 2015.}
\label{fig:base-prob-priors}
\end{figure}

\begin{figure}[htbp]
\centering
\includegraphics[width=1.0\textwidth]{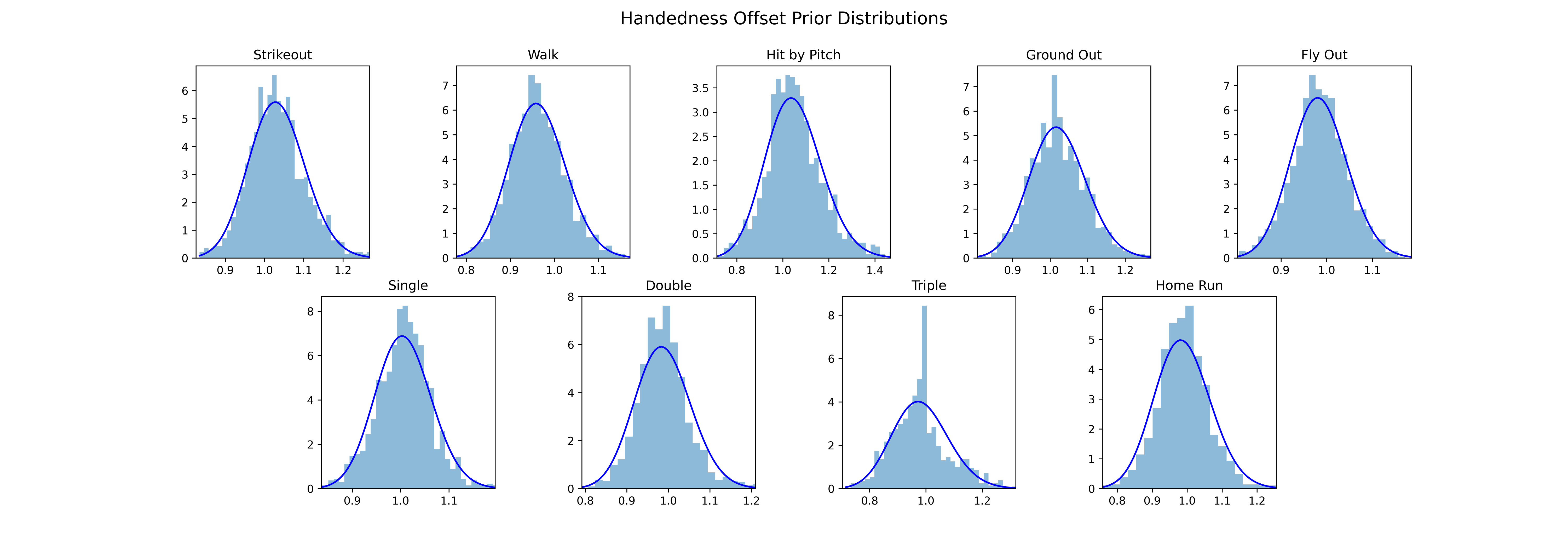}
\caption{Handedness offset prior distributions, capturing how pitcher performance differs between same-handed and opposite-handed matchups.}
\label{fig:handedness-offset-priors}
\end{figure}

\begin{figure}[htbp]
\centering
\includegraphics[width=1.0\textwidth]{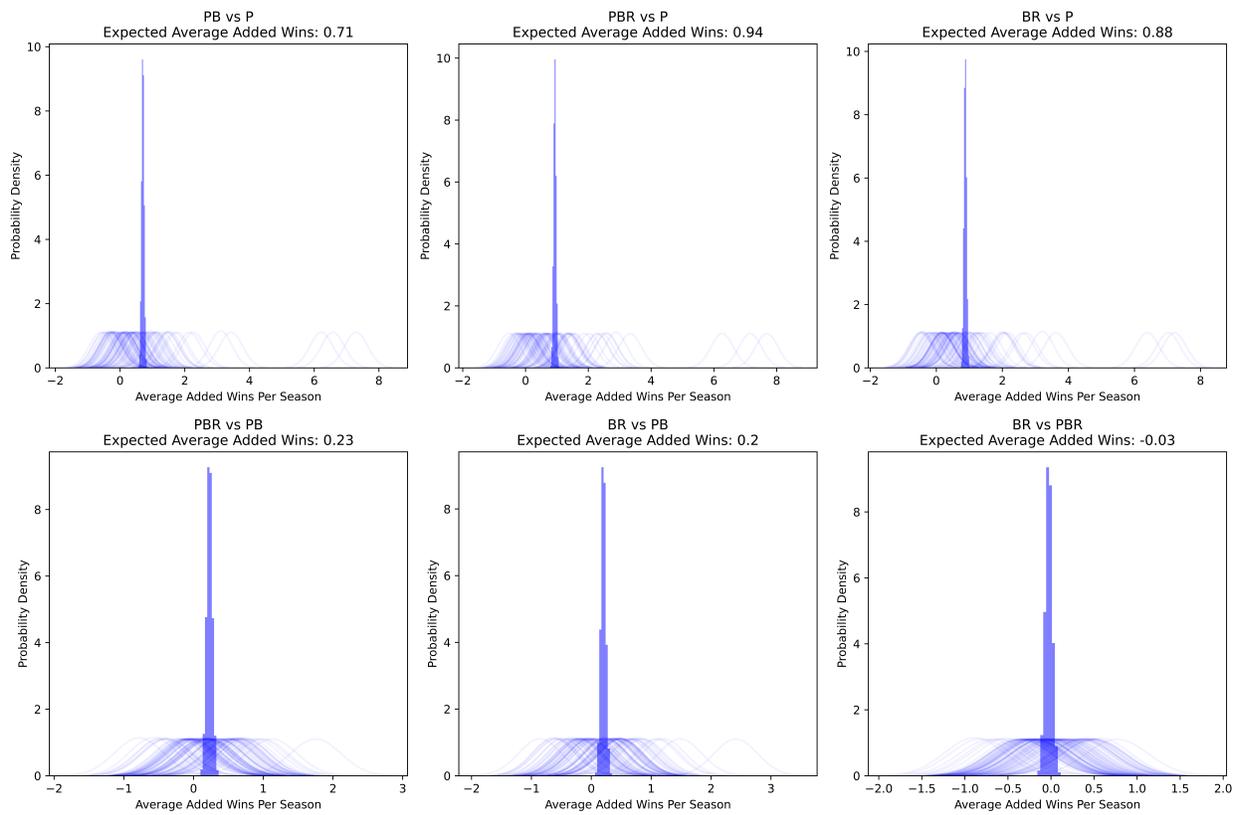}
\caption{Posterior distributions of Added Wins Per Season for each model, estimated via Monte Carlo simulation.}
\label{fig:posterior-added-wins}
\end{figure}

\begin{figure}[htbp]
\centering
\includegraphics[width=1.0\textwidth]{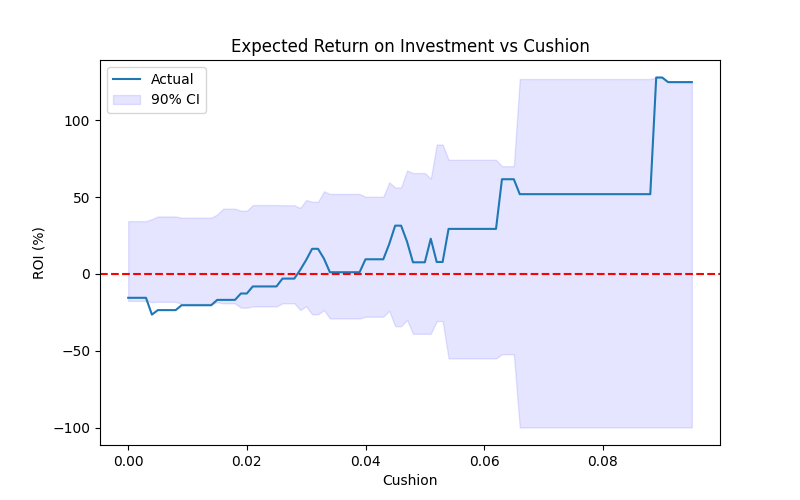}
\caption{Actual ROI that would have been made at different cushion thresholds, along with 90\% confidence intervals, assuming our model's predictions are correct.}
\label{fig:money-cushion-roi}
\end{figure}

\begin{figure}[htbp]
\centering
\includegraphics[width=1.0\textwidth]{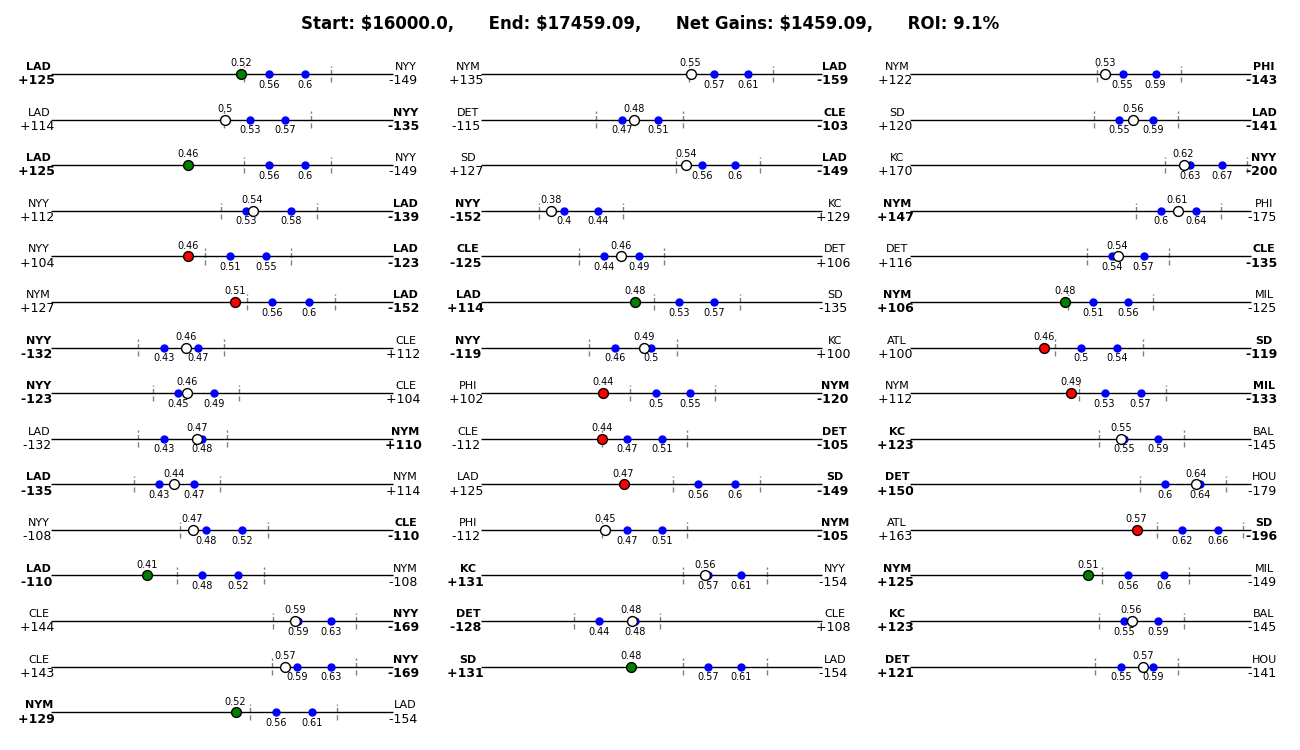}
\caption{Betting lines vs.\ our predicted win rates. Blue dots mark the overround, white dots mark unplaced bets, green dots represent winning bets, and red dots represent losing bets. Dotted lines denote cushion thresholds. Teams that actually won are shown in bold.}
\label{fig:betting-lines-plot}
\end{figure}

\noindent
Figure \ref{fig:base-prob-priors}. Base probability prior distributions for the 9 plate appearance outcomes, estimated from pitchers with more than 100 plate appearances since 2015. \dotfill \pageref{fig:base-prob-priors}\\
Figure \ref{fig:handedness-offset-priors}. Handedness offset prior distributions, capturing how pitcher performance differs between same-handed and opposite-handed matchups. \dotfill \pageref{fig:handedness-offset-priors}\\
Figure \ref{fig:posterior-added-wins}. Posterior distributions of Added Wins Per Season for each model, estimated via Monte Carlo simulation. \dotfill \pageref{fig:posterior-added-wins}\\
Figure \ref{fig:money-cushion-roi}. Actual ROI that would have been made at different cushion thresholds, along with 90\% confidence intervals, assuming our model's predictions are correct. \dotfill \pageref{fig:money-cushion-roi}\\
Figure \ref{fig:betting-lines-plot}. Betting lines vs.\ our predicted win rates. Blue dots mark the overround, white dots mark unplaced bets, green dots represent winning bets, and red dots represent losing bets. Dotted lines denote cushion thresholds. Teams that actually won are shown in bold. \dotfill \pageref{fig:betting-lines-plot}\\

\clearpage
\setcounter{page}{1} 

\section{Introduction}
Baseball is built on decisions---such as when to intentionally walk, whether to
pinch-hit, or when to pull a pitcher. These choices add up over the course of a
game, and teams have long sought ways to optimize them. Over the past few
decades, approaches to optimizing in-game strategies have ranged from simple
heuristics to highly sophisticated systems that incorporate game theory and a
broad array of player attributes.

We approach the optimization of in-game baseball strategy by modeling the game
as a back-and-forth sequence of management decisions---such as substitutions
and intentional walks---and plate appearances, which managers cannot directly
control. To optimize the game, one first must be able to model plate appearances
between any given batter and pitcher. This means, for any possible plate appearance,
one must be able to predict the probability of outcomes such as home runs,
strikeouts, walks, etc. We refer to any model that predicts plate appearance
outcomes along with the subsequent state transitions as a ``matchup model''
since it models plate appearance matchups between batters and pitchers. Once
one has a matchup model to represent the plate appearances in a game, they can
then use game theory to make optimal management decisions based on the
projections from the matchup model.

In prior work \cite{melville2025}, we mainly focused on the game theory aspect
of baseball optimization, using Tom Tango’s simple Marcel projections
\cite{tango2012} combined with generic, empirically calculated state transitions
as our matchup model. We then modeled a baseball game as a stochastic,
zero-sum, perfect-information, extensive-form game played between two managers
who each aim to maximize their team’s expected win probability. Although
finding true subgame perfect Nash equilibria for baseball is impossible due to
the enormous state space, we explored several alternatives that can approximate
optimal strategies. The best of these, an AI manager we call the “pseudo-full
game manager,” outperformed its competitors. We believe it comes close to
calculating true equilibria strategies---certainly far closer than any human
manager could.

However, a perfect equilibrium strategy is only truly optimal if the underlying
matchup model it relies on is also accurate. Although no model can perfectly
represent the highly stochastic interactions between batters and pitchers, in this
paper we explore several advanced models and measure their potential impact on
the game. We compare a range of approaches, from the most basic---treating all
batters the same and focusing only on pitcher tendencies---to more detailed
methods that incorporate batter-specific performance, recent trends, and base
running ability. Using the pseudo-full game manager, we then evaluate how each
model influences in-game decision-making and whether adding features offers
better strategic outcomes. Finally, we compare our best-performing model’s win
probability estimates to sportsbook betting lines to see how well our
game-theoretic approach aligns with real-world expectations.

\section{Matchup Models}
We define a \emph{matchup model} as a model that takes an MLB batter, pitcher,
and initial game state as input and outputs a distribution of game states that the
plate appearance could transition to. The game state is defined as the number of
outs and the location of any base runners (24 total states). Formally:
\begin{equation}\label{eq:state-transition}
\text{state}_{i+1} \sim f(\text{state}_{i}, \text{pitcher}, \text{batter}).
\end{equation}

We model these transitions in two parts. First, we predict an ``outcome
distribution,'' which is dependent only on the pitcher and batter, not on the game
state. The 9 possible outcomes are strikeout, walk, hit by pitch, ground out, fly
out, single, double, triple, and home run. Then, we predict a distribution of next
game states based on the current game state, outcome, and possibly the batter,
but not the pitcher. For instance, if a batter hits a single with two outs and a
runner on second base, we need to know the probabilities of the runner advancing
to third base, scoring, getting thrown out, and so on. We call these ``base running
distributions.'' When combined, the outcome distributions and base running
distributions form a complete matchup model. Our first three models focus
solely on outcome distributions and use the same empirically derived base
running distributions for all batters (the same ones we used in
\cite{melville2025}). Our fourth and final model creates custom base running
distributions for each MLB batter.

There are numerous ways to construct a matchup model. One common method used
in baseball is the \emph{log5 method}, which combines the batter, pitcher, and
league tendencies in log space in order to make outcome predictions
\cite{james1981}. Another approach is to train a neural network to predict plate
appearance outcomes based on player embeddings, as is done in \cite{damour2017}.
Currently, a trend in MLB analytics is to apply hierarchical Bayesian
models---and for good reason. Such models allow us to learn league-wide trends
from large datasets while refining individual player predictions, even for players
with limited plate appearances. By incorporating prior knowledge gleaned from
overall league behavior, these models balance observed performance for each
player with broader statistical tendencies, helping to avoid overfitting. All four
of the models discussed in this paper trace their foundations to a hierarchical
Bayesian variation of the log5 model, as described in \cite{doo2018}. However,
our models extend this framework to predict multinomial plate appearance
outcome distributions, rather than binomial distributions. This is done by
calculating the probabilities of each of the plate appearance outcomes separately
using the log5 method, and then renormalizing them, all within our Bayesian
framework.

\subsection{The Pitcher Only Model}
For the ``pitcher only'' model, we learn individual pitcher distributions for each
MLB pitcher in our dataset, while batter distributions are learned only at the
level of batting order positions (not specific batters)---hence the name
``pitcher only.'' Each pitcher and batting order has separate distributions for
same-handed matchups and opposite-handed matchups, but these distributions are
highly correlated with each other as part of our Bayesian hierarchy. The idea
here is that pitcher substitution decisions---often the most critical in a
game---are commonly guided by a pitcher’s attributes and the handedness of
upcoming batters, rather than the specific identity of those batters. We want to
see if this ``pitcher only'' model can advise near-optimal strategies even though
it does not know batter-specific information.

For pitcher $n$, the probability of the $i$th plate appearance outcome against a
random batter with handedness $h$ is denoted by $a_i(n,h)$:
\begin{equation}\label{eq:pitcher-dist}
a_i(n,h)=
\begin{cases}
\hat{a_i}(n)^{\,o_i(n)}, & \text{if } h=0 \\[6pt]
\hat{a_i}(n)^{\,1/o_i(n)}, & \text{if } h=1
\end{cases}
\end{equation}
where $o_i(n)$ gives the $i$th handedness offset for pitcher $n$,
$\hat{a_i}(n)$ is that pitcher’s base probability for outcome $i$, and $h$
denotes whether the batter is the same- or opposite-handed as the pitcher
(1 for same, 0 for opposite). All $o_i$ are sampled from a gamma prior
distribution, and all $\hat{a_i}$ are sampled from a beta prior distribution.
If the pitcher and batter are opposite-handed, then $a_i(n)$ is equal to
the base probability raised to the power of the handedness offset. If the
pitcher and batter are same-handed, then $a_i(n)$ is equal to the base
probability raised to the inverse power of the handedness offset. This
structure reflects that the probabilities for a given outcome, such as strikeouts,
are linked for same- and opposite-handed matchups rather than being independent.
As the handedness offset approaches 1, then both probabilities converge on the
base probability. If the handedness offset exceeds 1, the probability is higher
against same-handed batters, while values below 1 indicate higher probability
against opposite-handed batters. Below are the empirically estimated prior
distributions for baseline probabilities and handedness offsets for all 9 plate
appearance outcomes, based on MLB pitchers with more than 100 plate appearances
since 2015.

Similar to $a_i(n,h)$, $b_i(m,h)$ represents the probability of the $i$th outcome
for a batter in batting order $m$ against a random pitcher, and $c_i(h)$
represents the probability for the entire league. Both $b_i(m,h)$ and $c_i(h)$
are empirically calculated from 2015--2024 data, rather than being learned as
part of the Bayesian network. For a given plate appearance between pitcher $n$
and a batter in the $m$th batting order, all three probabilities are converted to
log ratios and combined as follows:
\begin{equation}\label{eq:S_i}
S_i(n,m,h) = P_i \cdot \ln\Bigl(\frac{a_i(n,h)}{1-a_i(n,h)}\Bigr)
 + B_i \cdot \ln\Bigl(\frac{b_i(m,h)}{1-b_i(m,h)}\Bigr)
 - (P_i + B_i - 1)\cdot \ln\Bigl(\frac{c_i(h)}{1-c_i(h)}\Bigr)
\end{equation}
The linear combination of these ratios in log space, $S_i(n,m,h)$, follows the
general log5 approach used in \cite{doo2018}. The weights $P_i$ and $B_i$ are
set for the entire league and sampled from a uniform prior distribution between
0.25 and 1.75, subject to the constraint that $1 \le (P_i + B_i) \le 2$. The
league weight is equal to $-(P_i + B_i -1)$, ensuring that the three weights sum
to 1. As explained in \cite{doo2018}, this negative league weight captures the
idea that if an extreme batter meets an extreme pitcher (both above or below
league average), the result will be even more extreme than either’s average.
When $P_i + B_i$ is near 1, this ``extreme factor'' is negligible; when $P_i +
B_i$ is close to 2, it becomes significant.

The log-space weighted average is then converted back to an unnormalized
probability $x_i(n,m,h)$:
\begin{equation}\label{eq:x_i}
x_i(n,m,h) = \frac{1}{1 + e^{-S_i(n,m,h)}}
\end{equation}
This same process is done for all 9 plate appearance outcomes, and the final
outcome distribution $X$ is given by:
\begin{equation}\label{eq:categorical}
X(n,m,h) \sim \text{Categorical}\Bigl(\frac{x_i(n,m,h)}{\sum_{j=0}^{8} x_j(n,m,h)}\Bigr),
\quad i \in \{0,1,\ldots,8\}
\end{equation}
We use these equations to construct a Bayesian inference network in which the
posteriors for pitcher rate and offset values, along with the league-wide pitcher
and batter log weights, are sampled via the NUTS algorithm in Python
\cite{pymcADVI,pymcNUTS,hoffman2014}. The final model then uses the posterior
means to predict plate appearance outcomes, and we combine those outcome
predictions with league-average base running transition probabilities to complete
the ``pitcher only'' matchup model.

\subsection{The Pitcher + Batter Model}
The ``pitcher + batter'' model follows the same structure as ``pitcher only,''
but it also learns rate and offset values for each MLB batter by sampling from
empirically estimated prior distributions. This creates an interdependent
Bayesian framework, in which each observed outcome is dependent on both pitcher
\emph{and} batter tendencies, where weights are still learned for the entire
league. This allows us to finetune estimates even for players with small sample
sizes or opponents that are unusually good or bad, such as a pitcher in a tough
division.

\subsection{The Pitcher + Batter + Recency Model}
Whereas the previous models assign equal weight to all samples from 2015 to
2024, the ``pitcher + batter + recency'' model gives more weight to recent plate
appearances when sampling posteriors. Specifically, we split our training data
into four chains, each used to sample separate posteriors. For each batter or
pitcher, their most recent 500 plate appearances appear in all four chains; the
next 500 appear in three of the four chains; the following 500 appear in two
chains; and the final 500 appear in only one chain. We exclude any plate
appearances beyond the most recent 2,000 for each player. This yields a
weighting of approximately 40\% for the most recent data, 30\% for the
second-most recent, 20\% for the third group, and 10\% for the oldest set. This
split was inspired by the Marcel projections, which weight the previous three
seasons at approximately 42\%, 33\%, and 25\% respectively
\cite{tango2012}. After sampling from each chain, we average the posterior means
to arrive at the model’s final rate estimates. We opt for multiple chains so that
all plate appearances factor in while still emphasizing newer data.

\subsection{The Pitcher + Batter + Recency + Base Running Model}
The ``pitcher + batter + recency + base running'' model predicts plate appearance
outcomes in the same way as the previous one but replaces league-average base
running transitions with player-specific transitions. We stratify players by
their base-stealing tendencies. For each MLB batter, we estimate a stolen-base
rate:
\begin{equation}\label{eq:posterior-rate}
\text{posterior predictive rate} 
= \frac{\alpha + x}{\alpha + \beta + n}
\end{equation}
where $x$ is the number of observed steals, $n$ is the total steal opportunities
(a pitch where the player is on first or second base with the next base open),
and $\alpha$ and $\beta$ come from a beta prior, fit to data from players with at
least 200 opportunities. We apply each batter’s posterior predictive stolen-base
rate to every base running transition in our dataset, then sort those transition
data into five evenly sized groups to empirically estimate base running
transition probabilities. We use these large groupings because there are 24
possible base-out states and 9 possible plate appearance outcomes, creating 216
transitions that must be estimated. To avoid sample-size issues, each batter’s
transition distribution is constructed by taking a weighted average of the five
group distributions, where the weights are determined by integrating over that
batter’s posterior stolen-base rate. Because we use the posteriors to determine
the weights, batters with large sample sizes are more likely to be placed solely
in one group while batters with smaller sample sizes are more likely to be more
evenly spread across all five groups, a desirable quality.

\subsection{Model Evaluation}
All four matchup models use the same training and validation datasets created
using the \texttt{pybaseball} statcast API \cite{pybaseball}. The training set
includes all MLB plate appearances from March 2015 through June 2024, while the
validation set comprises those from July 2024 through October 2024. For each
model, we generated predictions on the validation set and calculated cross
entropy loss (log loss) for both outcomes and transitions. We also computed the
geometric mean probability (GMP), defined as $e^{-\text{log loss}}$, which
provides a more intuitive sense of how well the model’s predicted probabilities
match actual results. To determine how much these models can improve teams’
chances of winning, we tested them in Monte Carlo simulations. Because our
``pitcher + batter + recency + base running'' model (BR) achieved the
lowest log loss, we used it to sample transitions during simulation. For
comparison, we recalculated both outcome and transition GMP for each model,
assuming the BR model as the ground truth.

\begin{table}[htbp]
\centering
\scriptsize
\begin{adjustbox}{max width=\textwidth}
\begin{tabular}{lcccccc}
    \toprule
    \textbf{Model} & 
    \textbf{Outcome Loss} & 
    \textbf{Outcome GMP} & 
    \textbf{Transition Loss} & 
    \textbf{Transition GMP} &
    \textbf{Outcome GMP (vs BR)} &
    \textbf{Transition GMP (vs BR)} \\
    \midrule
    P Model & 1.788 & 16.73\% & 1.171 & 31.00\% & 16.48\% & 30.81\% \\
    PB Model & 1.772 & 17.00\% & 1.168 & 31.09\% & 16.79\% & 30.94\% \\
    PBR Model & 1.771 & 17.02\% & 1.168 & 31.10\% & 16.82\% & 30.96\% \\
    BR Model & 1.771 & 17.02\% & 1.166 & 31.15\% & 16.82\% & 31.00\% \\
    \bottomrule
\end{tabular}
\end{adjustbox}
\caption{Comparison of log loss and geometric mean probability (GMP) on the validation set, as well as GMP recalculated assuming BR is ground truth.}
\label{tab:logloss}
\end{table}

Because of the highly stochastic nature of MLB plate appearances, the best model
(BR) actually performs slightly better on the validation set than it does when it
is itself ground truth. This may indicate that the Bayesian priors are too
restrictive, causing the model to be less confident than it should be. While this
may be refined in future work, the accuracy is satisfactory for the scope of this
paper. This is because we are more interested in simulating the \emph{relative}
difference between models and observing the in-game impact of these differences.

\section{Revisiting the 2024 MLB Playoffs}
To test how well our models could strategize in actual game scenarios, we
replayed each of the 43 games in the 2024 MLB playoffs 1.6 million times. For
each game, we used all four models to ``solve'' the game by approximating
subgame perfect Nash equilibria strategies, following the process outlined in
\cite{melville2025}. Solving a game first requires specifying batting orders,
starting pitchers, and substitution constraints. We obtained actual batting
orders and starting pitchers using the MLB Stats API \cite{mlbStatsApi} and set
each team’s available pitchers and batters based on who actually played in the
real contest. For teams that had fewer than six pitchers in a given game, we
added pitchers we believed were rested and healthy. Naturally, only the real MLB
managers know for certain which pitchers were truly available, but we consider
our approach a reasonable approximation.

\subsection{Simulation Results}
For each game in the 2024 playoffs, we first computed how many times each model
won when playing against itself as both the home team and the away team. For
each model, we then used a uniform prior to form a beta posterior distribution
of its true win rate against itself. Then, we replaced the opponent with each of
the three other models and recalculated the beta posteriors in the same way. The
difference between each of these posterior distributions and the original
distribution reflects how many additional wins the second model provides, in
that particular game, relative to the original model. Although the difference
between two beta distributions lacks a closed-form solution, we approximate it
by treating each beta distribution as a normal distribution. We repeat this for
every game, then sample uniformly across these normal distributions to form a
final posterior distribution describing how much better or worse the second
model is than the original. Finally, we normalize these results (both at the
individual-game and combined levels) to represent wins added over a 162-game MLB
season. Below we show both the game-specific and overall posterior difference
distributions for each combination of models.

In our simulations, the ``pitcher only'' model clearly performed the worst,
losing nearly a full game of relative win probability compared to the better
models. The ``pitcher + batter'' model was closer to the top models, but the gap
was still statistically significant. Meanwhile, the models factoring in recency
and base running were almost indistinguishable. Although the PBR model did
marginally outperform the BR model in simulation, we suspect this is just noise,
given that BR served as the ground truth. As a result, we view the two models as
essentially tied, especially since 0.0 falls comfortably within their posterior
difference distribution.

Interestingly, the value gap between these models does not match their
differences in log loss. Even though adding recency barely affected the log
loss, it had a notable impact on win probability (with PBR clearly outperforming
PB). Conversely, while base running produced a bigger change in log loss, it
appeared to have virtually no impact on win probability.

Based on our simulation, an MLB team could gain roughly one additional win simply
by improving its matchup projections by a fraction of a percent in GMP (the GMP
difference between PBR and P was only 0.0015), assuming all other aspects of
play remain optimal. Considering that an MLB position player contributing one
extra win in win probability added would place them among the top 50 players in
a typical season \cite{fangraphs2024}, enhancing the matchup model seems like a
more accessible and cost-effective strategy.

\subsection{Leveraging Simulations for Prediction}
After running our extensive playoff simulations for 2024, we thought it would be
interesting to compare those results with what actually happened. If we assume
that our base running model is correct and that MLB managers behave optimally
(two strong assumptions, admittedly), we can treat the simulated home win
rate---when both teams are managed by the base running model---as the expected
home win probability.

We compared our model’s win probabilities to state-of-the-art models by examining
the consensus betting lines from the 2024 playoffs \cite{oddsportal}. To assess
its predictive value, we calculated the theoretical profit or loss if our model
had wagered \$1000 on every game where its estimated win probability differed
enough from the betting market’s implied probability to exceed the \emph{overround}
(the margin sportsbooks rely on to ensure profitability). Our model’s win
probabilities fell within the overround margin for 15 of the 43 games, meaning
no bet was placed. For the remaining 28 games, where our model identified a
perceived edge, placing bets across all of them still resulted in an overall
loss. However, by introducing a ``cushion''---a stricter threshold beyond the
overround for placing bets---our model was able to turn a profit. Specifically,
with any cushion of at least 3\%, it would have produced positive returns on the
2024 MLB playoffs.

\begin{table}[htbp]
\centering
\scriptsize
\begin{adjustbox}{max width=\textwidth}
\begin{tabular}{cccccc}
\toprule
\textbf{Cushion} & \textbf{Bets Placed} & \textbf{Total Bets} & 
\textbf{ROI Lower Confidence} & \textbf{ROI Upper Confidence} & \textbf{Actual ROI} \\
\midrule
0\%   & 28 & \$28K & -18\% & +34\% & -15.5\% \\
1.5\% & 23 & \$23K & -18\% & +38\% & -16.9\% \\
3\%   & 16 & \$16K & -21\% & +48\% & +9.1\% \\
4.5\% & 10 & \$10K & -34\% & +56\% & +13.5\% \\
6\%   & 5  & \$5K  & -55\% & +74\% & +29.4\% \\
7.5\% & 3  & \$3K  & -100\% & +127\% & +52.0\% \\
9\%   & 2  & \$2K  & -100\% & +128\% & +128.0\% \\
9.5\% & 1  & \$1K  & -100\% & +125\% & +125\% \\
\bottomrule
\end{tabular}
\end{adjustbox}
\caption{Monte Carlo-sampled ROIs with various ``cushion'' thresholds when betting against consensus lines.}
\label{tab:ROI}
\end{table}

Assuming our win probability estimates are accurate, we then employed Monte
Carlo sampling to generate 90\% confidence intervals for return on investment
(ROI) under each possible cushion level.

We also provide a plot illustrating the betting lines in comparison to our
predicted win rates, along with a 3\% cushion. Interestingly, our model never
bet on the home team---likely because it does not factor in home field advantage,
even though home teams do win more often in reality. Accounting for home field
advantage could potentially be a valuable addition to future matchup models.

\section{Conclusion}
In this paper, we have shown that applying \emph{even modest refinements} to a
baseball matchup model can produce significant advantages. By moving from
basic approaches, such as our ``pitcher-only'' model, to more nuanced Bayesian methods, teams can capture extra wins and gain significant strategic
edges. Our simulations indicate that adding recency plays a surprisingly
important role in shaping in-game decisions even though it may not show up very much in the cross-entropy loss. However, while integrating base running
data provides a more robust description of player transitions, getting these transitions correct did not seem to play any role in decision making and win probability. Additionally, by comparing the simulated results with the real betting lines, we found that the forecasts of our best model align well with the predictions of the external market, suggesting a broader applicability for both the in-game strategy and predictive analytics. Future work could incorporate additional factors such as home field advantage and pitcher fatigue for even greater realism. Nevertheless, our results underscore that
small gains in projection accuracy, combined with game-theoretic decision-making,
can offer a large payoff in the pursuit of in-game success.

\section*{Acknowledgments}
We would like to thank Henry Mott for his invaluable assistance in assembling
realistic rosters for our simulation. Thanks to his familiarity with baseball, he
balanced bench and bullpen players so that each playoff team’s available players reflected authentic in-season usage. This level of detail added fidelity to our simulated games and strengthened our findings.

\bibliography{ref}

\begin{thebibliography}{10}

\bibitem{fangraphs2024}
Major league leaderboards.
\newblock
  \url{https://www.fangraphs.com/leaders/major-league?pos=all&stats=bat&lg=all&qual=y&type=3&season=2024&month=0&season1=2024&ind=0&sortcol=2&sortdir=default&pagenum=1&pageitems=2000000000},
  2024.
\newblock Accessed 2024-11-01.

\bibitem{pymcADVI}
Pymc advi documentation.
\newblock
  \url{https://www.pymc.io/projects/docs/en/latest/api/generated/pymc.ADVI.html},
  2024.
\newblock Accessed 2024-11-01.

\bibitem{pymcNUTS}
Pymc nuts documentation.
\newblock
  \url{https://www.pymc.io/projects/docs/en/stable/api/generated/pymc.sampling.jax.sample_numpyro_nuts.html},
  2024.
\newblock Accessed 2024-11-01.

\bibitem{oddsportal}
{MLB 2024 Results, Scores \& Historical Odds}.
\newblock \url{https://www.oddsportal.com/baseball/usa/mlb-2024/results}, 2025.
\newblock Accessed 2025-03-04.

\bibitem{damour2017}
Alexander D'Amour, Shane Jensen, and Abraham Wyner.
\newblock (batter|pitcher)\textasciicircum2vec: Statistic-free talent modeling
  with neural player embeddings.
\newblock In {\em MIT Sloan Sports Analytics Conference}, 2017.
\newblock Accessed 2024-11-01.

\bibitem{doo2018}
Woojin Doo and Heeyoung Kim.
\newblock Modeling the probability of a batter/pitcher matchup event: A
  bayesian approach.
\newblock {\em PLOS ONE}, 2018.
\newblock Accessed 2024-11-01.

\bibitem{hoffman2014}
Matthew~D. Hoffman and Andrew Gelman.
\newblock The no-u-turn sampler: Adaptively setting path lengths in hamiltonian
  monte carlo.
\newblock {\em Journal of Machine Learning Research}, 15:1593--1623, 2014.

\bibitem{james1981}
Bill James.
\newblock Log5 method for matchup probabilities.
\newblock Baseball Abstract, 1981.

\bibitem{pybaseball}
James LeDoux and \texttt{pybaseball} contributors.
\newblock {\texttt{pybaseball}}: Python package for mlbam statcast data.
\newblock \url{https://github.com/jldbc/pybaseball}, 2024.
\newblock Accessed 2024-11-01.

\bibitem{melville2025}
William Melville, Tristan Mott, Christopher Archibald, and David Grimsman.
\newblock An extensive investigation of strategies in baseball.
\newblock
  \url{https://www.sloansportsconference.com/research-papers/an-extensive-investigation-of-strategies-in-baseball},
  2025.

\bibitem{mlbStatsApi}
Todd Roberts.
\newblock {MLB Stats API Python Wrapper}.
\newblock \url{https://github.com/toddrob99/MLB-StatsAPI}, 2024.
\newblock Accessed 2024-11-01.

\bibitem{tango2012}
Tom Tango.
\newblock Marcel 2012.
\newblock \url{https://www.tangotiger.net/archives/stud0346.shtml}, 2012.

\end{thebibliography}
\bibliographystyle{plain}

\end{document}